# VTCdb: A transcriptomics & co-expression database for the crop species *Vitis vinifera* (grapevine)

**Darren CJ Wong[1], Crystal Sweetman [1], Damian P Drew [1, 2], Christopher M Ford[1§]**

[1]School of Agriculture, Food and Wine, University of Adelaide, South Australia, 5064, Australia.
[2]Section for Plant Biochemistry, Department of Plant and Environmental Sciences, University of Copenhagen, Frederiksberg 1871, Denmark
[§]Corresponding author

Email addresses:
DCJW: darren.wong@adelaide.edu.au
CS: crystal.sweetman@adelaide.edu.au
DPD: damian.drew@adelaide.edu.au
CMF: christopher.ford@adelaide.edu.au

## Abstract

### Background

Gene expression datasets in model plants such as *Arabidopsis* have contributed to our understanding of gene function and how a single underlying biological process can be governed by a diverse network of genes. The accumulation of publicly available microarray data encompassing a wide range of biological and environmental conditions has enabled the development of additional capabilities including gene co-expression analysis (GCA). GCA is based on the understanding that genes encoding proteins involved in similar and/or related biological processes may exhibit comparable expression patterns over a range of experimental conditions, developmental stages and tissues. We present an open access database for the investigation of gene co-expression networks is less available for the cultivated grapevine, *Vitis vinifera*.

### Results

We have constructed a grapevine gene co-expression database, VTCdb (http://vtcdb.adelaide.edu.au/Home.aspx) that offers an online platform for transcriptional regulatory inference in the cultivated grapevine. Using a condition-independent approach, the grapevine co-expression network was constructed using 352 publicly available microarray datasets from diverse experimental series, profiling approximately 9000 genes (40% of the predicted grapevine transcriptome). Examples of applications available from the online platform include the option to query genes, modules and biological processes of interest, and interactive network visualisation or analysis via CytoscapeWeb. To demonstrate the utility of the database, we present examples encompassing two fundamental biological processes, photosynthesis and flavonoid metabolism, within which novel associations are identified, while the recovered sub-networks reconcile established plant metabolic functions.

### Conclusions

Together, we present valuable insights into grapevine transcriptional regulation by developing a network model applicable to researchers in their prioritisation of gene candidates for on-going study of biological processes related to grapevine development, metabolism and stress responses.

## Background

Berries of the cultivated grapevine *Vitis vinifera* are one of the most highly valued horticultural crops in the world, and were among the earliest domesticated fruit crops in human history. The global production of grapes in 2009 was 67 million tonnes, harvested over approximately 8 million hectares of land, making the grapevine the most widely cultivated of fruit species (http://faostat.fao.org). Quality attributes of grapes, including characteristics such as aroma, flavour, colour and texture, have a profound impact on the wine produced from them, and therefore on the value of the grape crop itself. An in-depth understanding of gene expression and the regulation of metabolic pathways controlling various aspects of grapevine development and berry metabolism could provide insights into the genetic factors influencing fruit quality, and ultimately inform future vineyard germplasm and cultural practices.

Functional genomics studies in plants have contributed a systems-level understanding of how genes function and how an underlying biological process of a plant is governed by the cooperation of a set of genes. Such studies have been facilitated by the sequencing of full plant genomes that enable global gene identification, and the application of high throughput technologies such as microarrays and RNA sequencing. For many species, these high throughput technologies have been used to investigate gene expression in different tissues, in tissues at different stages of development, or in plants subjected to diverse conditions, and the large datasets produced have aided our understanding in many biological questions of interest. A survey from gene expression data repositories including the Gene Expression Omnibus [1] and Arrayexpress [2] revealed that a large number of expression datasets have been generated from plants, especially *Arabidopsis thaliana*, *Glycine max* (soybean) and *Oryza sativa* (rice), and involve diverse experimental conditions. Although these gene expression datasets have been primarily generated within a particular experimental context, the accumulation of large numbers of expression profiles has offered additional capabilities. These include comparative genomics between plant species, screening and functional assignment of gene candidates, the discovery of novel DNA motifs, and the dissection of regulatory networks.

Gene co-expression analyses (GCA) have received much attention recently, based on the notion that genes involved in similar or related processes may exhibit similar expression patterns over a range of experimental conditions [3, 4]. This "guilt by association" principle has been initially applied to gain insights into co-expressed gene modules within an organism [5, 6], to assign novel gene functions previously not ascribed to any biological processes, and to understand the evolution of gene expression and diversity across species and kingdoms [7, 8]. Within a co-expression network, genes and similarity relationships (commonly represented by correlation coefficients) are often visualised as nodes and edges respectively. The connection of nodes by an edge indicates a similar expression profile of the two nodes according to a particular similarity metric. For a given set of genes, the collection of these nodes and edges forms a network. Visualisation of the co-expression network enables the identification and description of densely connected gene clusters, referred to as modules, and an assessment of biological relevance can be achieved by investigating the functions of genes within each module [3, 9].

Many graph clustering algorithms have been developed with the aim of extracting functional modules comprising densely connected groups of nodes (representing co-expressed genes). Such algorithms can be classified as density-based and local search algorithms, hierarchical clustering, and other optimization-based algorithms [10]. In addition to the model plant *Arabidopsis*, these algorithms have also been applied to study co-expression networks in important crop species such as rice, barley and soybean [11, 12], with databases developed to store inferred modules and provide a user-friendly resource for plant biologists. Examples of successful studies reported using the "guilt by association" principle include the identification of genes involved in cellulose biosynthesis [13] and glucosinolate metabolism and regulation in *Arabidopsis* [14].

In *V. vinifera*, the grapevine Affymetrix microarray (http://www.affymetrix.com) is by far the most commonly used microarray platform to interrogate the transcriptome. Sequences selected from Genbank, grapevine expressed sequence tags and the NCBI RefSeq transcripts were the main sources for design and annotation, with approximately one third of the transcriptome represented on the array. Substantial sets of expression data have been generated for grapevines using the grapevine Affymetrix array. The collection of microarray studies to date features experiments involving berry development of various cultivars [15, 16], tissue-specific gene expression [17], and effects of stress responses (abiotic and biotic) [18, 19], among others. While these experiments were conducted to investigate a specific biological question focussing on selected biological processes, collectively the integration of these datasets provides a starting point to perform meta-analyses in this economically important crop.

In this study, over 350 publicly available microarray datasets related to the *Vitis vinifera* L. transcriptome were carefully selected to construct a global co-expression network (GCN) of grapevines using a combination of correlation rank transformation and graph-clustering. For graph clustering, we compared several clustering algorithms with respect to the quality of network division and performance. Using the optimal clustering solution, we systematically characterised every module of the GCN by combining Gene ontology (GO) with expression data. Finally, a publicly available database, named VTCdb (http://vtcdb.adelaide.edu.au/Home.aspx) was made available to query and browse the associated GCN. We discuss selected examples where we have identified well characterized biochemical pathways, and suggest potential novel gene functions and processes which can be inferred. Together, we demonstrate the suitability of gene co-expression approaches to mine coherent network modules that will facilitate gene discovery and the elucidation of regulatory networks of the grapevine.

## Construction and content

### Data acquisition and processing

Publicly available grapevine Affymetrix microarray (http://www.affymetrix.com) datasets were retrieved from Arrayexpress [2] and Gene Expression Omnibus [1]. From a total of more than 500 available arrays, arrays that were used to profile sources other than *V. vinifera* (i.e. other *Vitis* spp. or plants) were not used in this analysis, leaving behind raw CEL files of 394 arrays in total. A survey of the underlying experimental conditions represented by the arrays was assembled into six subcategories, covering a broad range of treatments and plant development stages such as tissue development, abiotic stress and hormone treatments (Table 1). This provided a broad basis for inferring gene co-expression relationships in grapevine.
CEL files were processed using RMAExpress (http://rmaexpress.bmbolstad.com/), using the default settings to compute robust multiarray average expression values. Potential outlier arrays were removed by visual inspection of raw perfect match data and iteratively discarding arrays that failed the quality control test (where expression values deviated significantly from the relative log expression and the normalized unscaled standard error). A total of 352 arrays were retained for subsequent analysis. Additionally, probesets with no hits or that did not satisfy a cut-off (e-value $\geq 1x10^{-5}$) when mapped against the latest predicted *V. vinifera* transcriptome (NCBI RefSeq) were also excluded. As a result, 14,210 probesets were retained, the robust multiarray average values were log2 transformed and this expression matrix data was used for gene co-expression analysis. Although the grapevine Affymetrix array can be used to measure the expression of 16,200 transcripts, only 14,210 probes satisfied our selection criteria with regards to having unique and single hits in the current *V. vinifera* RefSeq collection and were used as an input for construction of the grapevine co-expression network. Publicly available grapevine gene and probeset annotation (release 32) information was merged from Vitisnet [20] and NetAffx (http://www.affymetrix.com), respectively. In brief, these annotations contain unique gene identifiers for probesets across different databases, tentative functional and orthology information inferred from BLAST, ontology and pathway level information and subcellular localisation predicted using WoLF Psort [21]. To provide users with an alternative gene identifier to query the database, we assigned unique RefSeq identifiers to the probesets using the BLASTn module within PERL, by querying the consensus sequences on which probesets were designed against the latest *V. vinifera* NCBI RefSeq reference. For ontology annotations used in this study, previously assigned GO and Mapman annotations to probesets were achieved using B2G-FAR [22] and Mecator [23] pipelines, respectively. Correlations between all mapped probesets were calculated using Pearson's correlation coefficient (PCC) as a measure of similarity between expression profiles. Additionally, the mutual co-expression relationships between all gene pairs were also calculated (without applying any PCC cut-offs) by first transforming PCC into highest reciprocal ranks (HRR) [24]. Rank-based networks have been shown to be robust and offer greater advantages over correlation-based networks [25]. Such an approach has been frequently applied to retain weak but significant co-expression relationships and circumvent the unequal distribution of gene correlations for some genes when applying a fixed similarity threshold [25, 26]. This index of co-expression (PCC and HRR) serves as a basis for ranking co-expressed genes when using the 'guide gene' approach. Although estimation of the statistical significance of HRR [11] showed that HRR values ≤ 710 were significant ($P < 0.05$), we decided to display co-expressed genes with HRR ≤ 300 ($P < 0.02$) in the CoexQuery result page due to storage and retrieval constraints (see next section).

For co-expression analysis based on the capacity to identify modules of densely connected nodes, clustering algorithms such as HCCA [24], K-means [27], MCL [28], and MCODE [29] were tested. Algorithms were investigated for their quality of network division (modularity) and performance (specificity, sensitivity and F-measure) [10]. A HRR ≤ 30 ($P < 0.02$) co-expression network and expression matrix were used as an input for graph (HCCA, MCL, MCODE) and K-means clustering, respectively. To assist with the categorisation of partitioned modules according to their potential function or processes, we assessed the modules for enrichment primarily for Gene Ontology (GO) terms using g:Profiler [30] and GO-Module [31]. Enrichment using a different annotation scheme, Mapman [23], which is tailored primarily for plant-specific pathways and processes (independent of gene ontology) was also tested using PageMan [32]. Annotation terms (GO and Mapman) were considered significant if the corrected *P*-value was < 0.01 and there were at least two genes associated with the same annotation using the Fisher exact test and adjusted using false discovery rate (FDR) for multiple hypothesis correction. Of the five algorithms tested, HCCA was determined to partition the co-expression network efficiently and into biologically relevant modules (with better performance as measured by F-measure) in which genes involved in shared biological processes were successfully recovered (Additional File 1). Visualization of nodes and edges and their attributes were accomplished using Cytoscape [33]. In addition, expression specificities of individual probesets and modules were determined using the Std2Gx procedure [12]. Expression specificity (Std2Gx) expressed as values above one represent genes that are highly specifically expressed in the corresponding experiment compared with other genes and array samples. To determine the expression specificity of modules, the expression specificity values of all module members > 1 for a single assay (and repeated for all 352 assays) were counted and divided by the total number of module members. This gives a value (in %) showing the proportion of module members specifically expressed in a particular tissue/condition.

**Content and implementation of VTCdb**

VTCdb (http://vtcdb.adelaide.edu.au/Home.aspx) can be accessed using a user-friendly web interface, which includes tools to query, browse and visualise the co-expression network genes and modules. VTCdb runs on an Internet Information Server (IIS, version 7.0) web server containing data tables stored in MS Access (Microsoft Corporation Inc., USA). The web pages were built using a combination of ASP.NET 4.0, Javascript and Jquery 1.4. All calculations were performed in Python 2.7.3 prior to data deposition and database construction.VTCdb home page contains several search forms to retrieve co-expressed genes and related information using the co-expressed genes search, keyword search, enriched term search, and the option to browse meta-network interfaces (Figure 1).

For example, when a probeset ID is used as a query (Figure 2A), users will be directed to a page listing co-expressed genes (Figure 2C). Alternatively, the *keyword query* tool enables a broad search across all fields within the database and reports matching probesets with query terms (Figure 2B). The co-expressed genes page contains detailed information on query and target genes, a tabular list of ranked co-expressed genes that are colour coded and in ascending order of HRR. Sorting of any columns of the table can be done by clicking the respective headers of the table; this provides flexibility for the user interested in ranking the co-expressed genes list according to other indexes of co-expression (i.e. PCC). Clicking 'coexquery' in any row opens a new co-expressed genes result page for the chosen gene. Additional information such as co-expressed gene network visualisation (HRR≤30), predicted protein sub-cellular localisation using WoLF Psort, links to the ortholog co-expression network of the query gene in *Arabidopsis* (ATTED-II; http://atted.jp) and the associated module (when identifiable using HCCA) page to which the query gene belongs (Figure 3C) are contained within the co-expressed gene results page. The information provided is useful when an understanding of the conserved co-expressed genes between plant species, and reconciliation of co-expressed genes are of interest. The lists of co-expressed genes can be downloaded in a table format or sent to g:Profiler [30] for further functional analysis such as GO enrichment analysis.

Users are able to browse the modules obtained using HCCA directly from the CytoscapeWeb interface, or search functional annotation terms or identifiers of interest using enrichment keyword search (Figure 3A). Users will be redirected to the enriched term result page containing links to modules enriched with the query term (Figure 3B). Clicking the nodes representing modules within the meta-network provides details of the module size, number of edges and enriched GO and Mapman terms (Figure 3A) while double-clicking takes the user to the module results page, with detailed information on the module of interest (Figure 3C). The page lists three separate tables containing genes belonging to the module, combined GO and Mapman enriched terms (sorted by type and FDR significance), and expression specificity of the inferred module describing corresponding tissue/sample conditions of the majority of the probesets demonstrating specific expression. Additionally, network visualisation and analysis of the gene co-expression network (module) and enriched GO terms are provided using the CytoscapeWeb application. Functions such as displaying node/edge annotations, highlighting first-neighbours of nodes, and visualisation at different cut-off parameters are made available; these provide manipulation of the co-expression and enrichment network according to the users' preferences. All tables and networks can be downloaded individually or by bulk at the download page.

**Utility and Discussion**

To demonstrate the applicability and robustness of VTCdb web server for co-expression studies, we describe several applications using VTCdb query tools to investigate well-characterised biological processes and highlight gene co-expression networks that may be of biological interest in future grapevine research.

**Example application I: photosynthesis and fruit quality**

GOLDEN2-like transcription factors (GLKs) are essential in photosynthesis and chloroplast development [34] and play a key role in fruit quality determination [35]. The Arabidopsis pseudo-response regulator 2 (*APRR2*) is considered a gene related to but distinct from GLKs [36], although the functions of *APRR2* are not fully understood. Recently, functional analysis demonstrated that tomato and pepper *APRR2* homologs, when over-expressed *in planta*, increased plastid size, chlorophyll content and pigmentation [37], reminiscent of phenotypic features of recently published *GLK2* over-expression in tomato [35]. To provide useful clues that could be used to further elucidate the molecular mechanism of *APRR2* in grapevine, we searched the database using the keyword query "APRR2". In the keyword query result page, a unique probeset match, annotated as a putative grapevine APRR2 (XM_002279114.2; 1608397_at) was retrieved (Additional file 2: Figure S1). Clicking the probeset ID (with default HRR cut-off of 300) redirects to the result page and returns a lists of 173 co-expressed genes with an average HRR and PCC values of 194 and 0.8, respectively (Additional file 2: Table S1). This co-expressed gene list was predominantly occupied by a suite of genes encoding proteins associated with photosynthesis (Photosystem I & II subunits, light harvesting & core complexes and reaction center genes). Interestingly, we found that genes encoding isoforms of GDP-L-galactose phosphorylase (1613017_s_at; XM_002278303.2) and beta-carotene hydroxylase 2 (1617541_s_at; XM_002273545.1) involved in ascorbic acid and carotenoid biosynthesis, respectively were the top five ranked genes. Transcription factors such as *GLK1* (1617694_at & 1614851_s_at; XM_002275194.2) were also found within this list. Transcript profiles of these genes have been shown to be highly abundant in young berries, with decreasing abundance as berry development progresses [38, 39]. The top 100 co-expressed genes with *APRR2* were used as input into g:Profiler. While GO terms within the GO:BP category photosynthesis, (FDR< 4.95e-26), were highly enriched, terms such as L-ascorbic acid metabolic process (FDR<1.75e-3) and pigment biosynthetic processes (FDR<2.82e-3) were also significantly enriched (Table 2A; Additional file 2: Table S2). Reconciling findings from previous functional studies [34], we were able to recover co-expressed genes and determine the biological targets potentially involved for the observed phenotype. *APRR2* may have a conserved role in regulating a myriad of photosynthesis-related genes in plants [37]; our data, derived from its co-expression signature, suggest that this may hold also for grapevine. Furthermore, novel roles such as the regulation of oxidative stress responsive pathways can be inferred from the co-expressed gene sub-network of grapevine *APRR2*. As such, manipulation of grapevine *APRR2* could provide an opportunity for grape growers to breed for desirable agronomic traits such high vitamin A and C.

When *a priori* knowledge of target genes is not known, searches using terms of interest enriched within predicted modules can be conducted. For example, querying the GO ID/term "GO:0009765/photosynthesis, light harvesting" using the enriched term query tool (Figure 3A) retrieves modules 30 and 56 in the enriched term results page (Figure 3B). Both of the modules were highly enriched for GO:BP term "photosynthesis" (FDR<2.60e-38 and 6.60e-16, respectively) and GO:CC term “thylakoid” (2.50e-30 and 4.70e-17, respectively) (Additional file 2: Table S3). This observation is consistent with the many co-expression studies previously conducted in *Arabidopsis*, where genes involved in photosynthesis were found to form well-defined co-expression modules [40]. This is likely because photosynthesis requires coordinated assembly of proteins into large supercomplexes with numerous protein-protein interactions, and therefore their corresponding genes are expected to be highly co-expressed [41]. Module 30 had 83 nodes (corresponding to 72 genes) and 570 edges densely connected with genes predominantly involved in photosynthesis and the carbohydrate metabolism represented in green and blue, respectively (Figure 4A). Module 56 contained 59 nodes (corresponding to 50 genes), of which 27 had predicted roles in photosynthesis (~40%), while the others were mainly involved in antioxidant detoxification and were predicted components of the chloroplast ascorbate-glutathione cycle (Additional file 2: Table S3). These included chloroplastic ascorbate peroxidase (1618209_at), peroxiredoxin (1614204_at), thioredoxin (1608019_s_at) and glutatione-S-tranferase (1613992_at, 1609324_at, 1620356_x_at) represented as orange nodes (Figure 4B). The expression of such genes with genes associated with photosynthesis may be expected due to their role in the management of redox homeostasis and detoxification of oxygen radicals produced in excess by photosystems I and II in the chloroplast thylakoids [42]. Other neighbouring genes of photosystem complexes within the network that were not directly associated with photosynthesis in module 56 were involved in the glycine/serine cleavage system and tetrapyrole biosynthesis. The expression levels of these genes within module 56 were high in leaves, and during early berry development and abiotic stress (i.e. high-light, drought and salinity), while expression levels were very low in post-veraison berries, seeds and callus samples (Figure 4C; additional file 2: Table S4). This is not unexpected, considering their predominating roles in photosynthesis, and association with the chloroplast and photosynthetic tissues. From these data, it seems likely that genes comprising module 56 are involved in the maintenance of redox regulation and homeostasis during photosynthesis and related processes.

**Example application II: flavonoid metabolism**

Natural products derived from the phenylpropanoid pathway are secondary metabolites such as flavonoids, which encompass *inter alia* the subclasses anthocyanins, flavonols and proanthocyanidins, and are widely distributed across the plant kingdom. In plants, flavonoids are known for their antioxidative capacities, fulfilling many diverse functions including protection against abiotic stress [43], plant-pathogen/herbivore interaction and plant development. In grapevines, MYBPA1 regulates several structural genes of the phenylpropanoid/flavonoid and downstream proanthocyanidin pathway [44]. Using MYBPA1 as a keyword query (Figure 2A) returned 3 matching probesets (1619579_at, 1616094_at, 1611091_s_at) (Figure 2B). Selecting 1616094_at as query returned 41 co-expressed genes with average HRR and PCC values of 112 and 0.6 respectively. As expected, the structural genes encoding chalcone synthase, chalcone isomerase, phenylalanine ammonia lyase, dihydroflavonol 4-reductase, flavanone 3-hydroxylase, leucoanthocyanidin dioxygenase of the flavonoid pathway were among the top co-expressed genes (Additional file 2: Table S5). GO enrichment analysis using g:Profiler revealed that GO:BP terms such as flavonoid biosynthetic process (GO:0009813;FDR<1.9e-14) were significantly enriched (Table 2B; Additional file 2: Table S6). This corroborated previous transcriptomic profiling and promoter studies [44, 45] and demonstrates the ability of gene co-expression to recover a large proportion of known downstream target genes.

Conversely, module 50 (searchable via the network browser interface or enrichment term query tool with terms “flavonoid”) (Additional file 2: Figure S2) contained 230 nodes and 809 edges, and included many genes associated with the general phenylpropanoid/flavonoid and shikimate pathways, sugar modification, transport, kinases, transcription factors and several corresponding to proteins of unknown function. GO enrichment analysis showed that many of the enriched GO:BP terms were associated with “phenylpropanoid and flavonoid” biosynthetic processes (FDR≤2.3e-11) (Additional file 2: Table S7). The main flavonoid biosynthetic pathway genes (chalcone synthase, chalcone isomerase, phenylalanine ammonia lyase, dihydroflavonol 4-reductase, flavanone 3-hydroxylase, leucoanthocyanidin dioxygenase), as well as some transcription factors and several transferase/transport proteins were densely connected, while several other genes involved in further upstream and downstream pathways were also within the module. The majority of genes from the flavonoid pathway (~90%) were represented on the array, and were often co-expressed with two MYB transcription factors, *MYBA1* (1615798_at) and *MYBPA1* (Figure 5A). In grapevines, the transcription factor *MYBA1* has been shown to transcriptionally activate UDP glucose:flavonoid 3-O-glucosyltransferase (1619788_at, 1617171_s_at), the final step of anthocyanin biosynthesis [46]. Here, nodes surrounding *MYBA1*, particularly UDP glucose:flavonoid 3-O-glucosyltransferase gene, were seen to share strong, direct co-expression links (HRR≤10) and indirect links with genes encoding glutathione S-transferase and Caffeoyl-CoA O-methyltransferase, further complementing the role of MYBA1 suggested previously in regulating synthesis and vacuole sequestration of anthocyanins in grapevines. Additionally, the majority of the genes (>90%) were specifically expressed in berry skins (during véraison through to ripening) (Figure 5B) coinciding with the tissues and developmental programming of flavonoid accumulation. While many of the genes implicated in various steps of the pathway have been identified, including regulatory genes, there are still many nodes annotated as proteins of unknown function (~23 nodes) (Figure 5A), that may qualify as candidates for biosynthetic or regulatory gene products of this large gene family based on co-expression relationships and expression patterns.
The regulation of genes associated with photosynthesis and flavonoid metabolism displays a conserved co-expression network structure across nine different plant species [11, 12]. Thus, the co-expression analysis performed here largely confirms previous work while revealing new and additional roles for some of the uncharacterised grapevine genes, and demonstrates the utility of the grapevine co-expression network generated in this study.

**Miscellaneous features**

Under the “additional tools” page, we have included a user interface to query metadata information of grapevine berry development (Additional file 3) [38]. This includes absolute gene expression level/profiles, clusters of differentially expressed genes, expression comparison between platforms/cultivars and search (when available) associated *de novo* contigs assembled using Trinity [47].

**Comparison to similar co-expression studies**

We note that the platform used in this study represents 40% of the predicted grapevine transcriptome (around 9000 genes), while more recent platforms such as the Nimblegen array have up to 90% transcriptome coverage. Therefore with our approach, certain genes and their underlying transcriptional profiles cannot be measured. However, the grapevine Affymetrix Genechip has been used to catalogue many more experimental conditions than other platforms, thus facilitating the discovery of various condition-related relationships during various abiotic stress, biotic stress, hormone treatments, light exposure and other experimental treatments. To our knowledge, Fasoli and co-workers [48] performed GCA using the grapevine Nimblegen array dataset covering a wide-range of healthy tissue only (162 array samples) to infer transcriptome relationships of processes such as photosynthesis and flavonoid metabolism as a subset of their study. Nevertheless, we demonstrate that the co-expression relationships obtained using photosynthesis and flavonoid pathway-related genes were highly similar. Furthermore, we were able to identify novel transcriptional regulatory mechanisms supported by combined network and functional analysis in plants [37, 44], providing an example whereby our GCN can be used to infer gene function. The predicted modules using graph clustering were of high biological relevance and may offer new biological insights (i.e stress responses) of many uncharacterised genes within these modules. In addition, network visualisation tools which provide users to visually analyse or dissect co-expression networks add to another dimension of GCA interpretation. Therefore, VTCdb offers a one-stop online platform for grapevine gene co-expression and transcriptomics analysis.

**Future developments**

With the rising trend of RNA-sequencing analysis in the study of transcriptional regulation [38] and increasing use of the grapevine Nimblegen array [39, 48, 49], co-expression network inference could lead to better discovery of gene modules due to increased sensitivity, accuracy and coverage of transcript profiles. Therefore, biannual updates of the database will be conducted when new arrays are present or sufficient arrays from other platform for GCA becomes available. Due to the large proportion of uncharacterised genes, we are in the process of functionally annotating them on the basis of gene co-expression analysis and expression patterns.

## Conclusions

Gene co-expression analysis of the grapevine transcriptome has enabled us to uncover additional relationships from publicly available microarray data. This meta-analysis approach has facilitated the comprehensive annotation of functions to unknown genes and the discovery of functional modules in grapevines. We envisage the utility and potential of VTCdb(http://vtcdb.adelaide.edu.au/home.aspx) to provide further valuable information in hypothesis-driven studies and to aid grapevine researchers in their prioritization of gene candidates for further study towards the understanding of biological processes related to many aspects of grapevine development and metabolism.

## Availability and requirements

All results discussed within this study and additional tools to query pre-constructed networks, perform additional gene co-expression, expression meta-analysis and annotation searches are available freely at VTCdb (http://vtcdb.adelaide.edu.au/home.aspx). VTCdb supports all major web-browsers, preferably Google Chrome and Mozilla FireFox for visualization and performance purposes.

## Abbreviations

VTCdb: Vitis Transcriptomics and Co-expression database; GCA: gene co-expression analysis; GCN: gene co-expression network; GO: gene ontology; PCC: Pearson's correlation coefficient; HRR: highest reciprocal rank; FDR: false discovery rate; GLK: GOLDEN2-like transcription factors; APRR2: Arabidopsis presudo-response regulator 2.

## Competing interests

The authors declare that they have no competing interests.

## Authors' contributions

DCJW conceived the study, compiled and analysed the microarray data, performed co-expression data analysis, constructed the database platform and drafted the manuscript. CMF, CS, DPD participated in co-expression data analysis, design and coordination of the study and assisted in drafting the manuscript. All authors read and approved the final manuscript.

## Acknowledgements

This work was part-supported by Australia's grape growers and winemakers through their investment body, the Grape and Wine Research and Development Corporation, with matching funds from the Australian Government (project UA 10/01). DCJW is supported by a postgraduate research scholarship from the University of Adelaide. DPD received funding from the European Union Seventh Framework Programme (*FP7/2007-2013*) under grant agreement 275422, which supported a Marie Curie International Outgoing Fellowship. The authors would like to thank David Contreras Pezoa for his assistance with the *de novo* transcriptome assembly data.

## Figures

**Figure 1 - Screenshots of VTCdb home page displaying different search forms**

(A) The home page contains a brief introduction into VTCdb webserver. To begin queries, select 'click to search genes or processes of interests' to cascade (B) co-expressed genes and (C) keyword search forms. Selecting 'click to browse network' will cascade (D) enriched term search and the (E) browse meta-network interface.

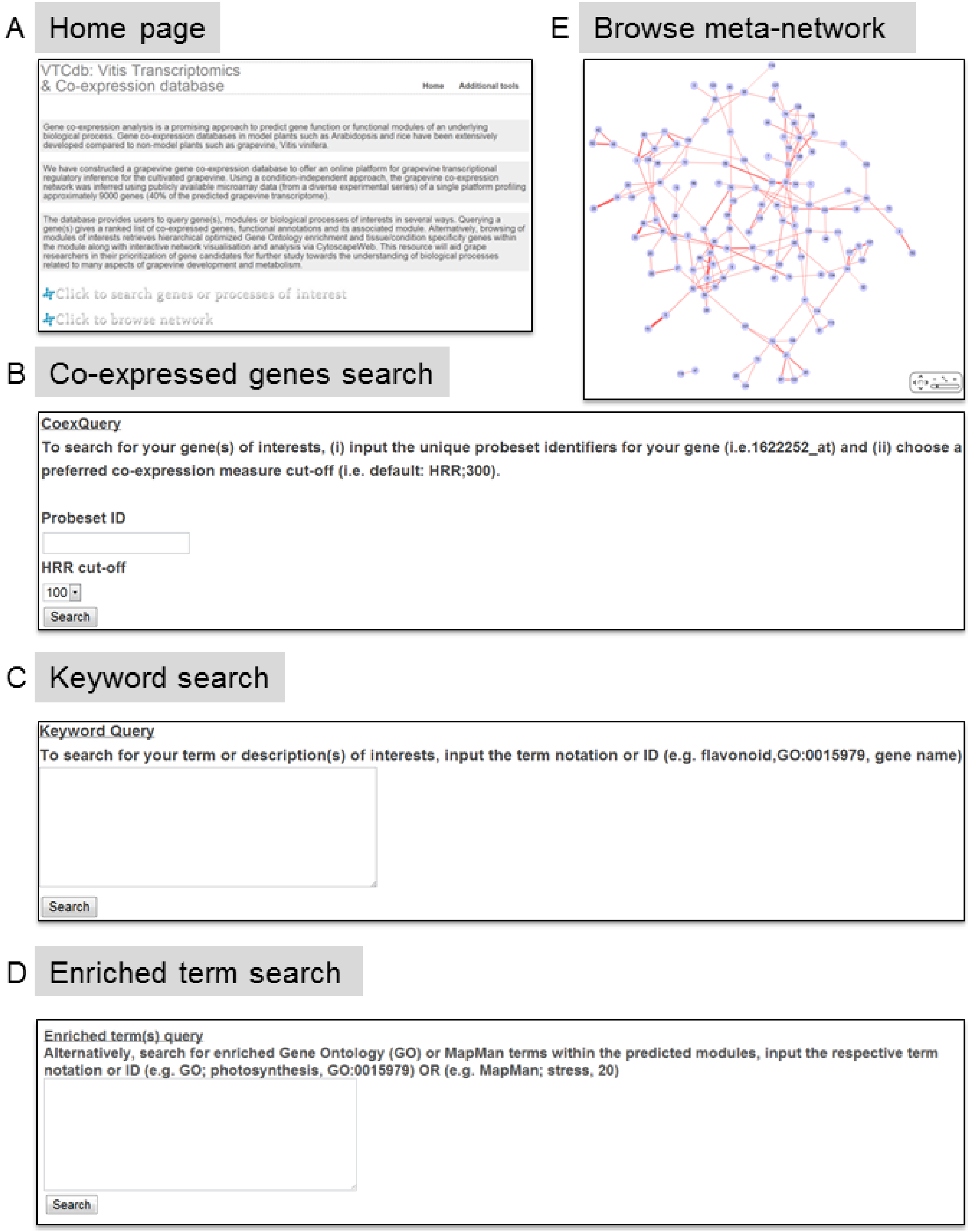

**Figure 2 – Screenshots of VTCdb pages relevant to the utility and output of probeset and keyword query searches.**

(A) To begin, input a probeset ID (e.g. 1616094_at) or a specific term (e.g. MYBPA1) of interests into the *co-expressed genes* and *keyword search* textbox, respectively. (B) Keyword results page containing probesets (links to co-expressed genes result page) and annotations with matching terms if keyword search is executed. (C) Co-expressed genes result page containing detailed information on query gene, co-expressed gene list, visualisation of network and tools to retrieve and/or submit to other webservers for functional enrichment analysis.

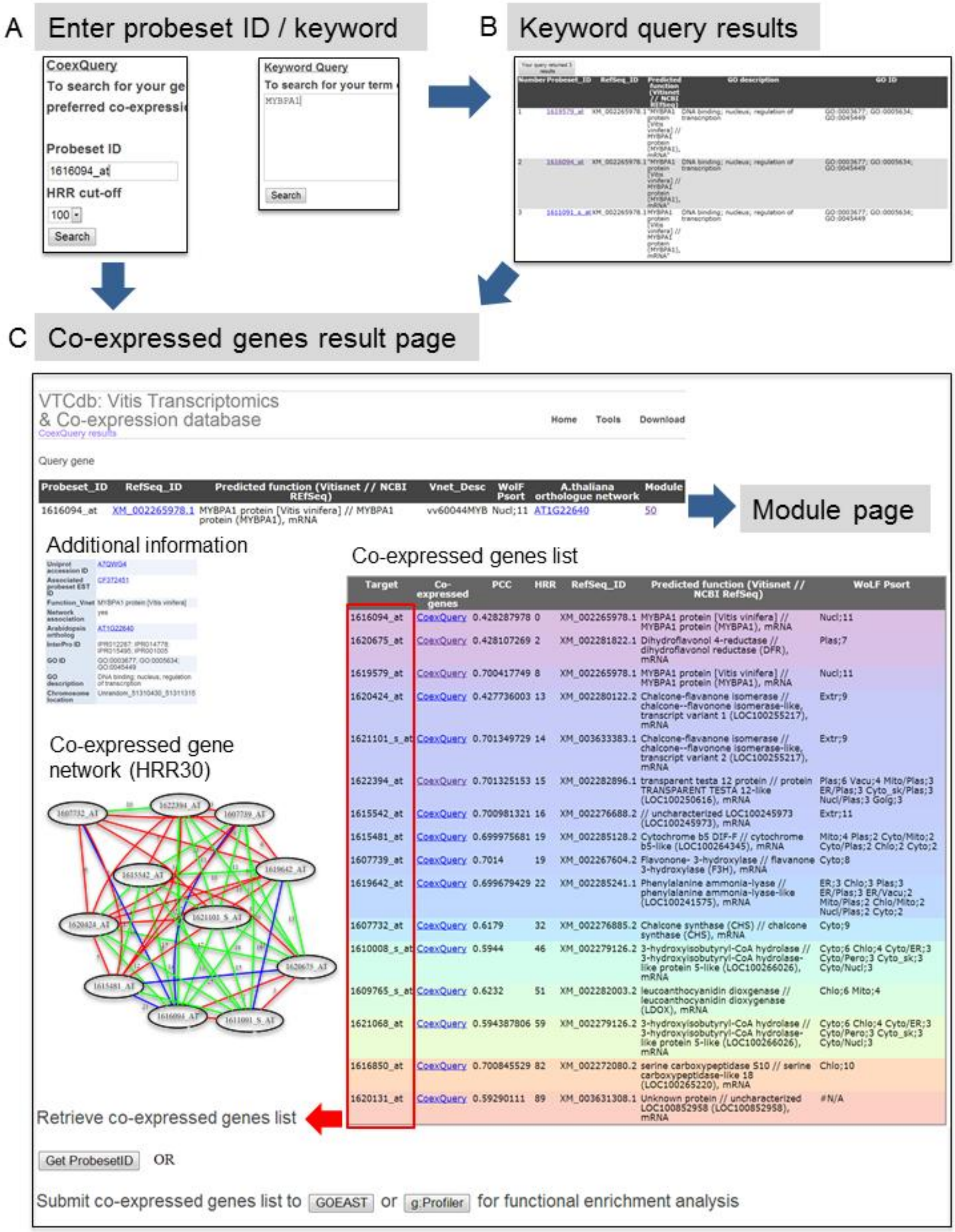

**Figure 3 - Screenshots of VTCdb pages relevant to the utility and output of the module browsing interface and enriched term search**

(A) The module browsing interface and enriched term query allow interactive display of inter-module network and search modules containing matching enriched terms, respectively. (B) Enriched term result page display matching module (and links to module page) enriched with query term of interests. (C) Module result page contains list genes, functionally enriched terms (when available), expression specificity of genes belonging to associated module and interactive visualisation (gene co-expression network and enriched GO term).

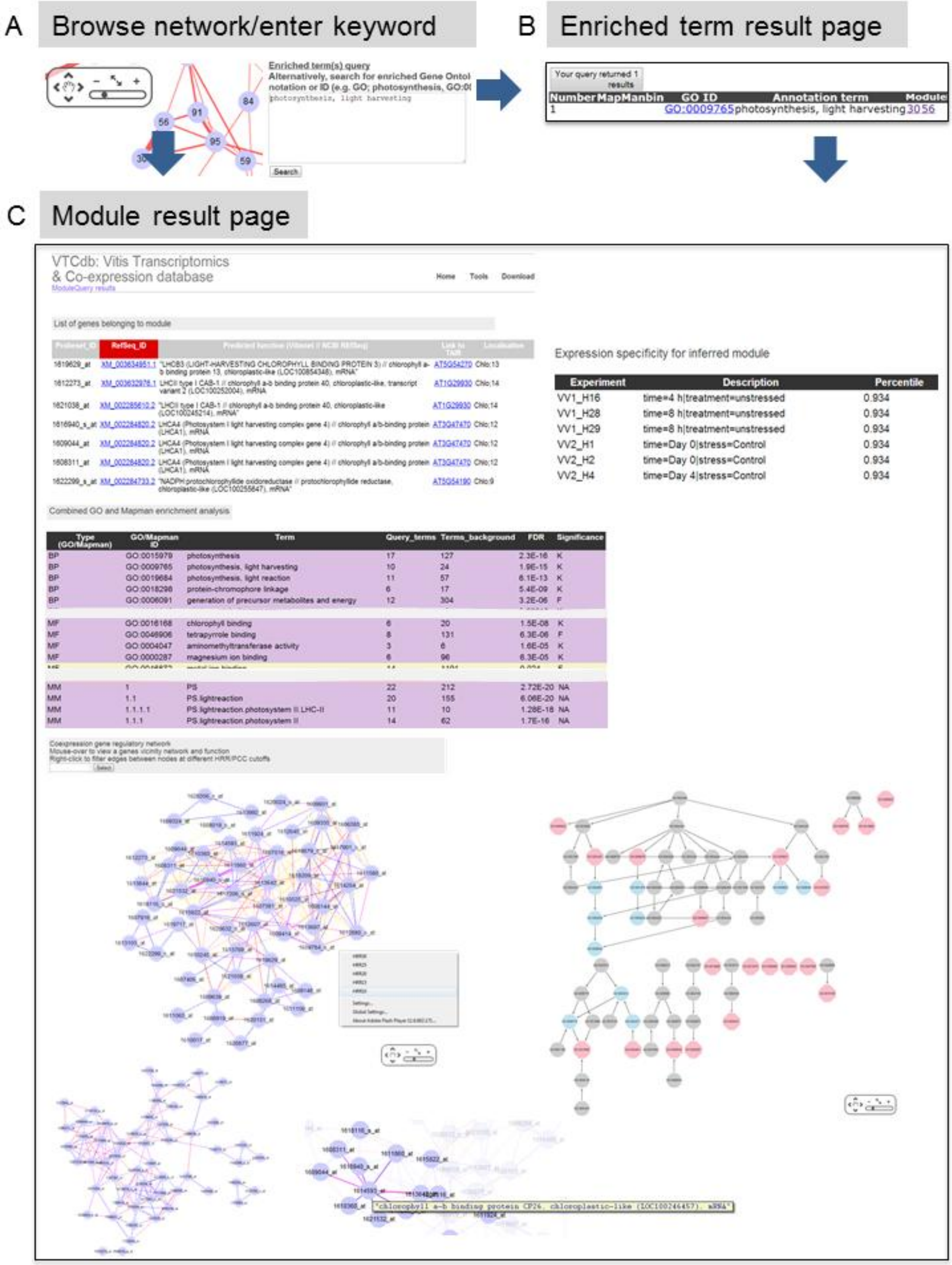

**Figure 4 – Predicted modules involved in photosynthesis**
(A) Module 30 contains 83 nodes and 570 edges densely connected with genes predominantly involved in photosynthesis and the carbohydrate metabolism represented in green and blue, respectively. (B) Module 56 contained 59 nodes (corresponding to 50 genes) involved in photosynthesis, oxidative stress management, tetrapyrole biosynthesis and glycine/serine cleavage system colored in green, yellow, blue and red, respectively. (C) Expression specificity of module 56. Green columns indicate the experiments in which the probesets (90% or greater) demonstrated specific expression (Std2Gx > 1.0).

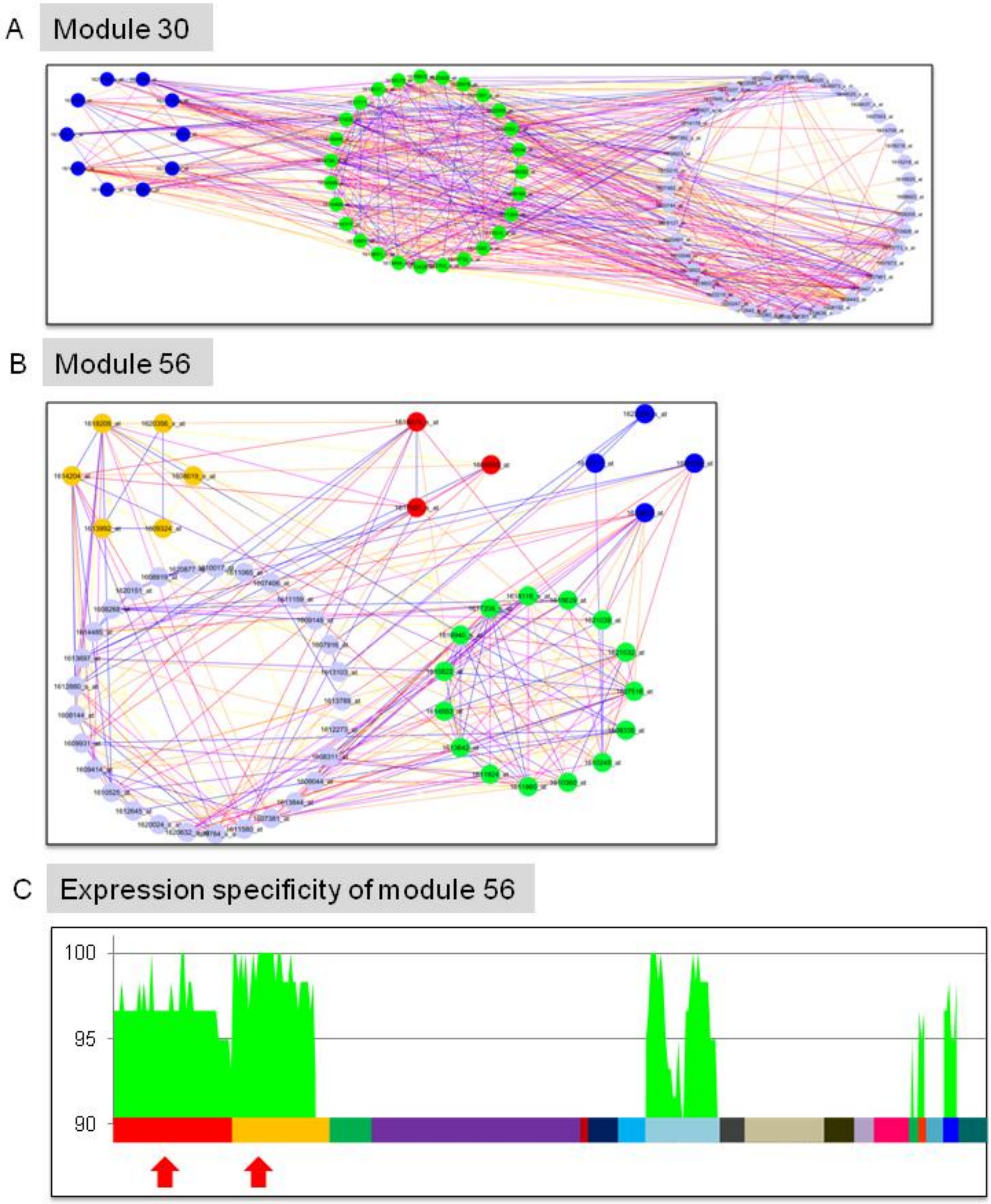

**Figure 5 – Predicted module involved in flavonoid, amino acid and related metabolism**
(A) Module 50 contains 230 nodes and 809 edges, and included many genes associated with the general flavonoid, aromatic amino acid, lignin and lipid pathways represented in purple, blue, red and orange respectively. Nodes coloured in green represent probesets whose function remains unknown or uncharacterised based on homology searches. (B) Expression specificity of module 50. Purple columns indicate the experiments in which the probesets (90% or greater) demonstrated specific expression (Std2Gx > 1.0).

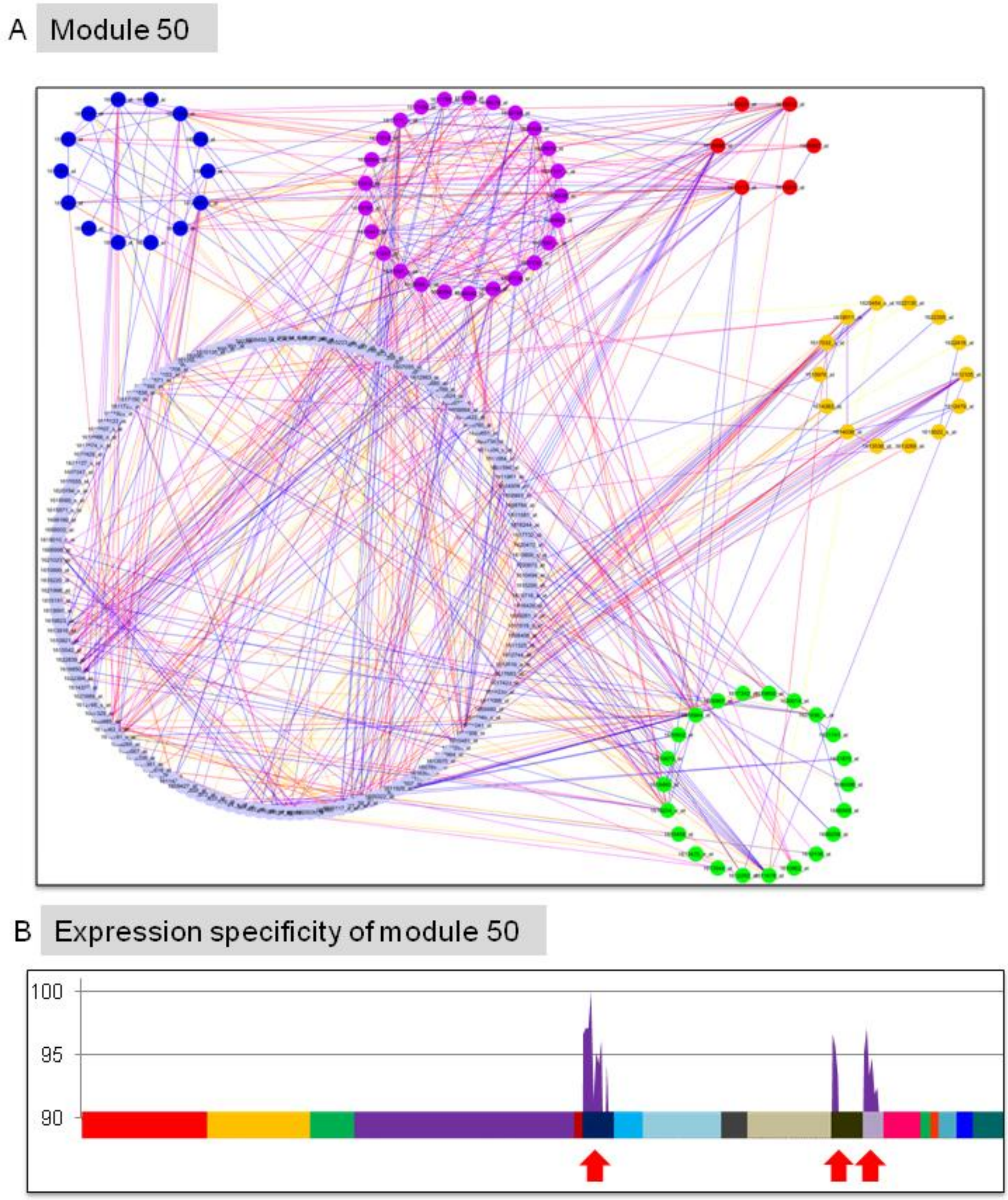

## Tables

**Table 1 - Summary of publicly available arrays analysed**

| Category | Number of experiments | Samples (CEL) | Fraction (%) |
|---|---|---|---|
| Abiotic stress | 5 | 89 | 22.6 |
| Biotic stress | 4 | 32 | 8.1 |
| Chemical treatment | 3 | 35 | 8.9 |
| Hormone treatment | 4 | 29 | 7.4 |
| Mutant/over-expressers | 2 | 10 | 2.5 |
| Tissue development (fruit) | 7 | 126 | 32.0 |
| Tissue development (non-fruit) | 7 | 73 | 18.5 |
| Total (unique) | 18 | 394 | |

**Table 2 - Summary of enriched gene ontology terms associate with *APRR2* and *MYBPA1***

(A) Prioritized GO terms of the top 100 co-expressed genes of grapevine *APRR2* and (B) MYBPA1. Listed are the type of GO domains, GO ID, GO term, the number of corresponding terms within the background, number of genes containing in query and the adjusted *P*-value (FDR). See Additional file 2 for full table listing.

(A)

| GO type | GO description | Term B/G | Query | | *P*-value |
|---|---|---|---|---|---|
| BP | photosynthesis | 149 | 20 | | 4.95E-26 |
| BP | response to blue light | 36 | 5 | | 7.95E-07 |
| BP | abaxial cell fate specification | 6 | 3 | | 1.05E-06 |
| BP | L-ascorbic acid metabolic process | 7 | 1 | | 1.75E-03 |
| BP | pigment biosynthetic process | 82 | 3 | | 2.82E-03 |
| BP | xanthophyll biosynthetic process | 6 | 1 | | 4.51E-03 |
| BP | negative regulation of abscisic acid mediated signaling pathway | 9 | 1 | | 4.51E-03 |
| BP | inorganic anion transport | 67 | 2 | | 5.41E-03 |
| BP | defense response by callose deposition in cell wall | 25 | 1 | | 6.27E-03 |
| BP | response to hormone stimulus | 669 | 7 | | 7.13E-03 |
| BP | carotene metabolic process | 12 | 1 | | 9.02E-03 |
| MF | GDP-galactose:glucose-1-phosphate guanylyltransferase activity | 1 | 1 | | 2.51E-04 |
| MF | GDP-D-glucose phosphorylase activity | 1 | 1 | | 2.51E-04 |
| MF | low affinity phosphate transmembrane transporter activity | 1 | 1 | | 1.00E-03 |
| MF | hydroxycinnamoyltransferase activity | 1 | 1 | | 4.01E-03 |
| MF | inorganic anion transmembrane transporter activity | 75 | 2 | | 6.76E-03 |
| MF | quercetin 4'-O-glucosyltransferase activity | 28 | 1 | | 7.02E-03 |
| MF | glutamate dehydrogenase [NAD(P)+] activity | 4 | 1 | | 7.02E-03 |
| MF | magnesium-protoporphyrin IX monomethyl ester (oxidative) cyclase | 1 | 1 | | 9.28E-03 |
| | activity | | | | |
| CC | photosynthetic membrane | 311 | 21 | | 5.89E-21 |

(B)

| GO type | GO term | Term B/G | Query | | *P*-value |
|---|---|---|---|---|---|
| BP | flavonoid biosynthetic process | 55 | 7 | | 2.49E-14 |
| BP | response to UV-B | 47 | 4 | | 1.81E-07 |
| BP | cellular modified amino acid biosynthesis | 41 | 2 | | 2.07E-03 |
| BP | maintenance of seed dormancy | 4 | 1 | | 4.10E-03 |
| BP | response to jasmonic acid stimulus | 113 | 2 | | 9.49E-03 |
| CC | plant-type vacuole membrane | 57 | 3 | | 4.86E-05 |
| CC | endoplasmic reticulum | 372 | 4 | | 6.49E-03 |
| MF | naringenin-chalcone synthase activity | 3 | 2 | | 6.62E-06 |
| MF | chalcone isomerase activity | 3 | 2 | | 7.72E-06 |
| MF | dihydrokaempferol 4-reductase activity | 1 | 1 | | 5.13E-04 |
| MF | leucocyanidin oxygenase activity | 1 | 1 | | 2.82E-03 |

Additional files

Additional file 1

Performance and modularity evaluation of several widely used clustering algorithms used in this study. Figure 1 illustrates the (A) F-measure, (B) specificity, (C) sensitivity and (D) estimated modularity scores obtained from different clustering solutions. Performance assessment was evaluated by calculating the fraction of modules (Specificity) enriched with at least one annotation and the fraction of annotations (Sensitivity) enriched in at least one module at FDR < 0.01 (defining significantly enriched terms). The two terms were then summarized as F-measure, as the harmonic mean between specificity and sensitivity [(2 x Specificity x Sensitivity/ (Specificity + Sensitivity)]. To determine the quality of network division, the modularity values of the partitioned network were estimated. The modularity function aims to quantify how well a network is partitioned into modules with scores ranging between -1 and 1. A network with high modularity contains modules with dense intra-module edges but sparse inter-module edges. Often the higher the modularity is the better quality of network division, and scores > 0.3 usually indicate a good partition. To provide a fair comparison between the different algorithms, settings that cluster large proportion of nodes was used and predicted modules with fewer than 10 probes, which are often are biologically meaningless, were subtracted from the total modules predicted. For HCCA, parameters were adjusted so that the desired module size was between 40 and 200 using a network neighbourhood size of 3 while the inflation score was set to 1.2 for MCL. For K-means clustering, the desired number of clusters, *k* was set to 100 and number of iterations set to 1000 with probesets not assigned to any cluster by HCCA were also excluded from clustering. For MCODE the degree cutoff, node score cutoff, k-core, and maximum depth were 10, 1, 3 and 3, respectively.

| | GO:BP | | | GO:MF | | | GO:CC | | | Mapman | | | Modularity |
|---|---|---|---|---|---|---|---|---|---|---|---|---|---|
| | Specificity | Sensitivity | F-measure | Specificity | Sensitivity | F-measure | Specificity | Sensitivity | F-measure | Specificity | Sensitivity | F-measure | |
| HCCA (*n* =3) | 0.45652 | 0.22864 | 0.305 | 0.44203 | 0.12891 | 0.2 | 0.2971 | 0.43432 | 0.353 | 0.65942 | 0.15141 | 0.24627 | 0.66 |
| MCL (*I* =1.2) | 0.41333 | 0.19477 | 0.26477 | 0.30667 | 0.10059 | 0.15149 | 0.21333 | 0.2252 | 0.21911 | 0.5375 | 0.12833 | 0.2072 | 0.68 |
| MCODE (10,1,3,3) | 0.40171 | 0.18476 | 0.253 | 0.39316 | 0.10645 | 0.168 | 0.28205 | 0.24129 | 0.26 | 0.552 | 0.11968 | 0.19671 | 0.46 |
| Kmeans (*k* =100) | 0.40625 | 0.15396 | 0.223 | 0.375 | 0.08984 | 0.145 | 0.125 | 0.26471 | 0.17 | 0.44792 | 0.06633 | 0.11555 | NA |

Additional file 2: Additional file 2.xlsx, 214K
http://www.biomedcentral.com/imedia/1427290988952533/supp2.xlsx
Additional file 2 contains 10 worksheets of VTCdb pages (screenshots) relevant to enriched term query and browse module interface for example application 1 (Figure S1), a list of co-expressed genes (HRR ≤ 300, *P*-value< 0.002) for grapevine APRR2 (Table S1), a list of all GO terms enriched associated with the top 100 APRR2 co-expressed genes (Table S2), a table containing list of genes and associated information of module 30 and 56 (Table S3) and a list of all experiments and corresponding expression specificity values inferred from module 56 (Table S4). Table S5 contains the full list of co-expressed genes (HRR ≤ 300, *P*-value< 0.002) surrounding *Vv MYBPA1*(Table S5), a list of all GO terms enriched associated with the top 100 *VvMYBPA1* co-expressed genes (Table S6), VTCdb pages (screenshots) relevant to enriched term query and browse module interface for example application 2 (Figure S2), a table containing a list of genes and associated information of module 50 (Table S7) and a list of all experiments and corresponding expression specificity values inferred from module 50 (Table S8).

Additional file 3
Additional file 3 contains screenshots of VTCdb pages relevant to RefSeq query interface. (A) To begin input NCBI RefSeq identifier of interests or type the any terms of interests using the keyword query. (B) For the keyword query, a result page containing matching keyword terms will be displayed together with their corresponding information. (C) The RefSeqQuery page contains the information pertaining to the information on the RefSeq identifiers of interests, absolute gene expression level/profiles, clusters of differentially expressed genes, compare these profiles between different platforms and grapevine cultivars and retrieve associated *de novo* contigs (when available).

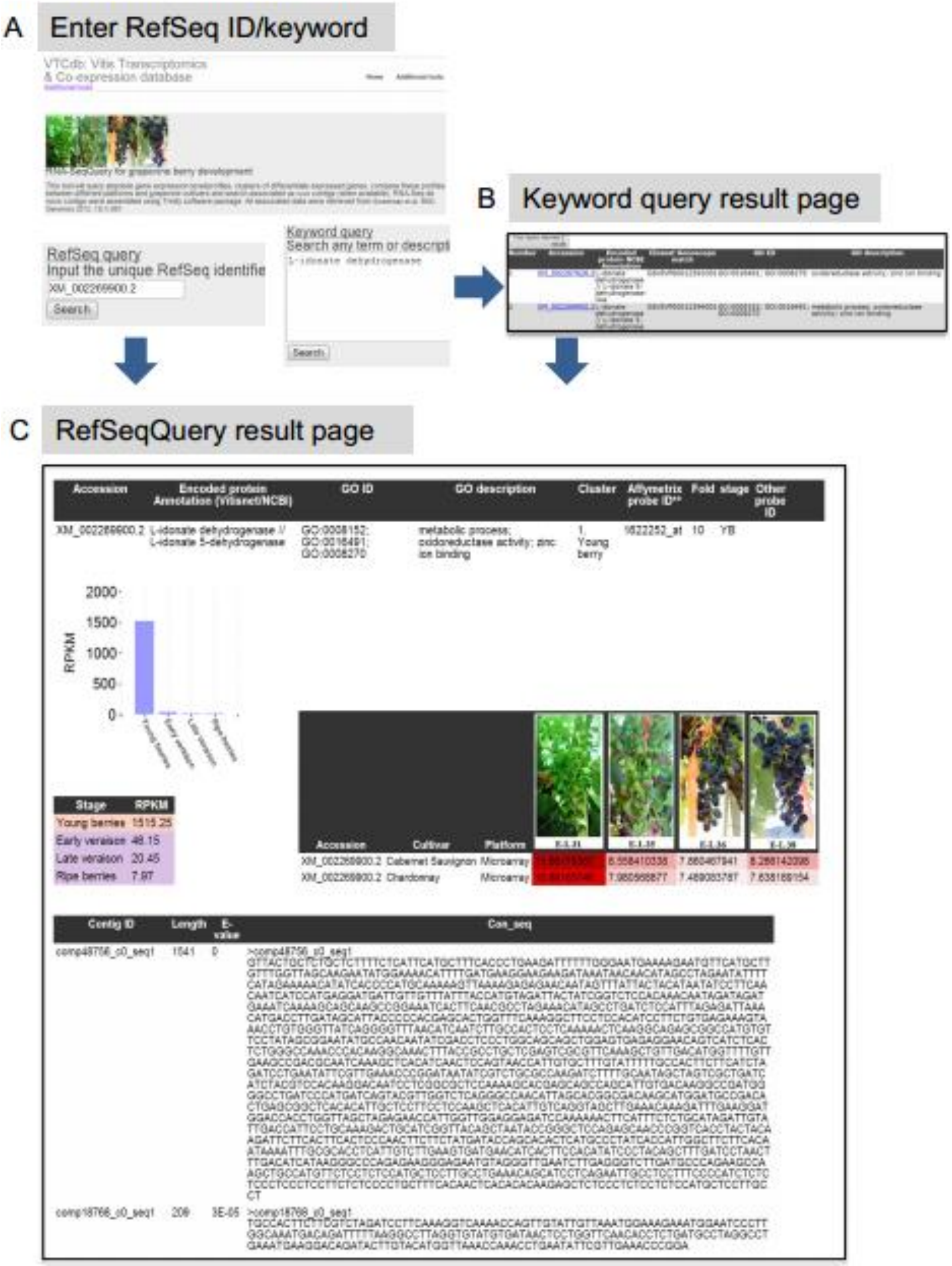
A Enter RefSeq ID/keyword
B Keyword query result page
C RefSeqQuery result page